\documentclass[twocolumn]{aastex62}

\usepackage{easyReview}
\usepackage{fontawesome}
\usepackage{hyperref}
\usepackage{amsmath}
\submitjournal{ApJ}
\accepted{April 12, 2019}
\shorttitle{\texttt{PICASO}}
\shortauthors{Batalha et al.}

\begin{document}

\title{Exoplanet Reflected Light Spectroscopy with \texttt{PICASO}}

\correspondingauthor{Natasha E. Batalha}
\email{nbatalha@ucsc.edu}

\author[0000-0003-1240-6844]{Natasha E. Batalha}
\affil{Department of Astronomy and Astrophysics, University of California–Santa Cruz, Santa Cruz, CA 95064, USA}

\author{Mark S. Marley}
\affiliation{NASA Ames Research Center, MS 245-3, Moffett Field, CA 94035, USA}

\author{Nikole K. Lewis}
\affiliation{Department of Astronomy and Carl Sagan Institute, Cornell University, 122 Sciences Drive, Ithaca, NY 14853, USA}

\author{Jonathan J. Fortney}
\affiliation{Department of Astronomy and Astrophysics, University of California–Santa Cruz, Santa Cruz, CA 95064, USA}



\begin{abstract}
Here we present the first open-source radiative transfer model for computing the reflected light of exoplanets at any phase geometry, called \texttt{PICASO}: Planetary Intensity Code for Atmospheric Scattering Observations. This code, written in Python, has heritage from a decades old, well-known Fortran model used for several studies of planetary objects within the Solar System and beyond. We have adopted it to include several methodologies for computing both direct and diffuse scattering phase functions, and have added several updates including the ability to compute Raman scattering spectral features. Here we benchmark \texttt{PICASO} against two independent codes and discuss the degree to which the model is sensitive to a user's specification for various phase functions. Then, we conduct a full information content study of the model across a wide parameter space in temperature, cloud profile, SNR and resolving power.
\end{abstract}



\section{Introduction} \label{sec:intro}
Across all the state-of-the-art pipelines that exist to study atmospheric composition and climate from exoplanets, about half a dozen have been developed for transiting science, a few of which are open-source \citep[e.g.][]{madhu2009temp, line2012info,benneke2012atmospheric,waldmann2015tau,barstow2017consistent,zhang2018forward}. This abundance of model development has overall improved the quality of all of these models, has contributed to interesting model inter-comparison studies \citep[e.g.][]{baudino2017toward}, and increased accessibility of traditionally private codes.

On the other hand, for observations of reflected light from directly imagined exoplanets there have only been two, neither of which are open source \citep{lupu2016developing, lacy2018wfirst}. An additional branch of models, used for Solar System/Earth science also exists to compute reflected light from planetary atmospheres (\texttt{NEMESIS}, \citet{irwin2008nemesis}; \texttt{DISORT}, \citet{stamnes1988disort}; \texttt{PSG}, \citet{villanueva2018psg}). \texttt{NEMESIS} is well-vetted, and has been used for retrieving composition from dozens of observations of Jupiter \citep[e.g.][]{irwin2019analysis}, Neptune \citep[e.g.][]{irwin2019probable}, Titan \citep[e.g.][]{thelen2019abundance} and many more. \texttt{DISORT} is an open-source forward model, however cannot be used to retrieve atmospheric composition. Note that all retrieval models consist of a versatile and fast forward model that can be wrapped in a statistical algorithm. \texttt{DISORT} contains several hard-wired assumptions for terrestrial conditions and is not versatile/fast enough to use in a retrieval framework. \texttt{PSG} is the retrieval tool of the ExoMars mission and has been used for other investigations such as Earth and the NASA Infrared Telescope Facility. Lastly, ray-tracing forward models, such as \citet{dyudina2016reflected}, would also be computationally intensive and complex to wrap into a retrieval framework for exoplanets.

With the detection and analysis of reflected light from optical phase curves \citep{demory2013kep7,esteves2015changing, niraula2018discovery} and optical photometry \citep{evans2013blue,barstow2014clouds, garcia2015probing, webber2015effect,lee2017dynamic}, and the onset of reflected light direct imaging missions on the horizon, such has \textit{WFIRST} and \textit{ELTs}, \citet{spergel2013wfirst}, there has been an increasing demand for an accessible, versatile reflected light code.

Here, we present the \textbf{P}lanetary \textbf{I}ntensity \textbf{C}ode for \textbf{A}tmospheric \textbf{S}cattering \textbf{O}bservations (\texttt{PICASO}). It is available through \texttt{Github}\footnote{\href{https://github.com/natashabatalha/picaso}{\faGithub:\texttt{Github}}}, and can be installed through \texttt{pip} or \texttt{conda}. Tutorials for running the code are available online\footnote{\href{https://natashabatalha.github.io/picaso}{\faCode:Code Tutorial}} along with an in-depth physics tutorial for the derivation of the radiative transfer of the code\footnote{\href{https://natashabatalha.github.io/picaso_dev}{\faSearch:Physics Tutorial}}.  

\subsection{The Heritage of the Code}
The methodology of \texttt{PICASO} partly originates from the \texttt{Fortran} albedo spectra model described in several studies of planetary objects within the Solar System \citep{mckay1989thermal, marley1999thermal} and beyond \citep{marley1999reflected}. These models utilized radiative transfer  methods
described in \citet{toon1977physical, toon1989rapid} and only included the capability to compute monochromatic scattered radiation observed at full phase.  Later, \citet{cahoy2010exoplanet} introduced the capability to compute the monochromatic scattered radiation observed \emph{at any phase angle}. Since then the model has been widely used in several studies of exoplanets including: retrievals of exoplanet atmospheres \citep{lupu2016developing, nayak2017atmospheric} sulfur hazes in giant exoplanet atmospheres \citep{gao2017sulfur},  Earth analogues in reflected light \citep{feng2018earth}, water absorption in cool giants \citep{macdonald2018exploring}, and color classification of directly imaged exoplanets \citep{batalha2018color}, among others. 

While the code bifurcated across several of these analyses (e.g. updates to molecular opacities, various ways to regrid the atmosphere, varying phase functions, different sources of scattering and cloud opacity) the bulk of the radiative transfer in the code has remained relatively similar since the original publications of \citet{mckay1989thermal,marley1999reflected,cahoy2010exoplanet}. Individual changes for each analysis lack trace-ability since the age of version control through \texttt{Github} is a recent phenomenon. 

\texttt{PICASO} is the first open-source compilation of all these previous works. It is written in \texttt{Python} and is designed to be user-friendly and versatile enough to handle all the use cases that have come before it, and additional use cases yet to be explored. 

\subsection{Exoplanet Diversity \& the Need for Versatility}
\begin{figure*}
\centering
\includegraphics[width=0.9\textwidth]{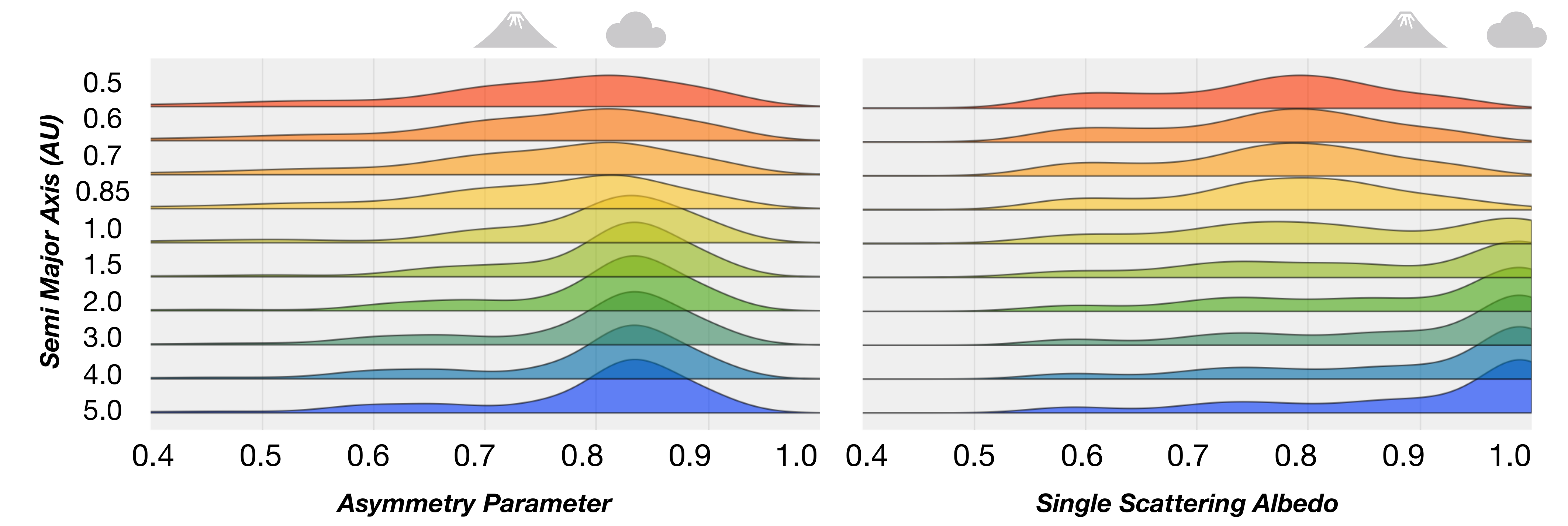}
\caption{Aggregated asymmetry parameters, and single scattering albedos from the grid of models used in \citet{batalha2018color}. Each distribution showcases the diversity of values we should expect for different temperature exoplanets ranging from semi-major axis$=0.5-5$ AU. For reference, the cloud icon represents the asymmetry and single scattering of a cirrus cloud, and the volcano icon represents the same of aged volcanic aerosols \citep{thomas2002radiative}. \textbf{Main Point:} Exoplanet atmospheres will exhibit a diverse range in optical properties.}
\label{fig:asy}
\end{figure*}
\begin{figure}
\centering
\includegraphics[width=\columnwidth]{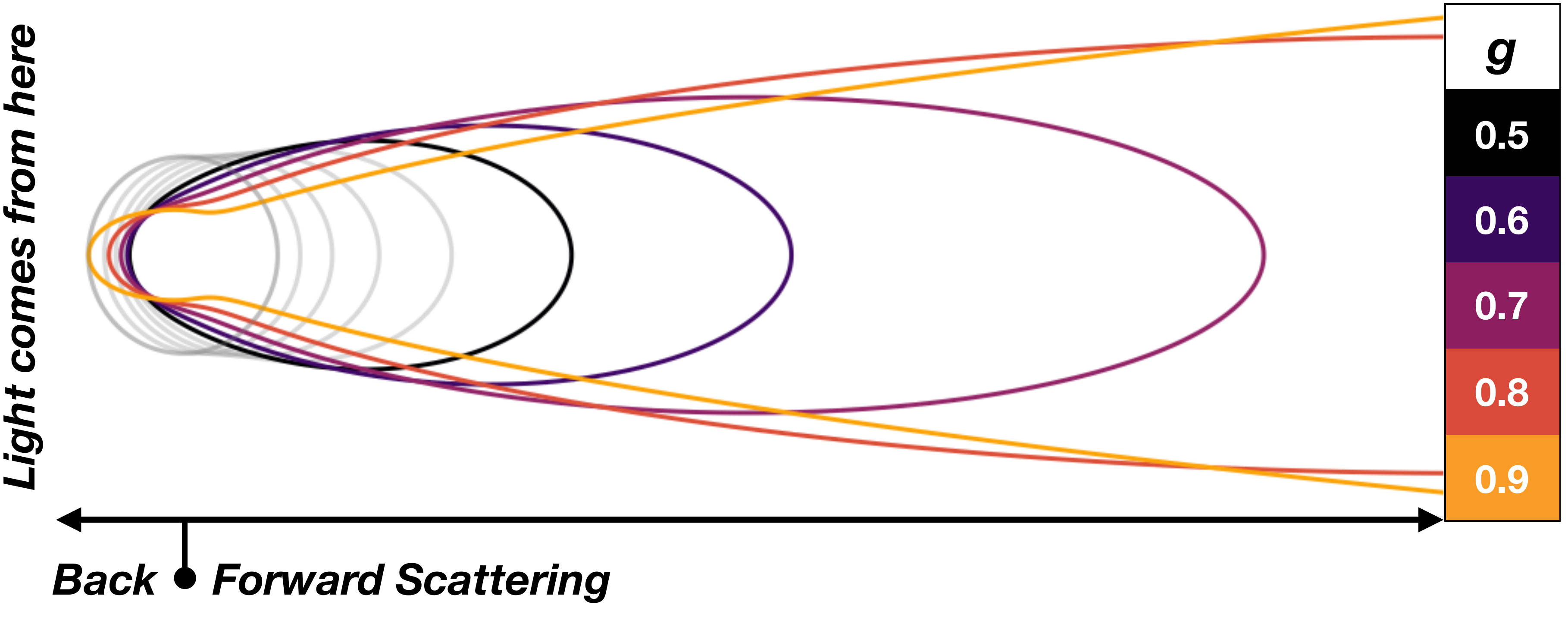}
\caption{A two-tern Henyey-Greenstein phase function for a range of asymmetry values.\textbf{Main Point:} Seemingly small changes to $g$ propagate to large differences in the phase function behavior. \href{https://natashabatalha.github.io/picaso_dev\#slide03}{\faSearch}}
\label{fig:hg}
\end{figure}

At the foundation of any reflected light code is an assumption of a scattering phase function, $p(\cos\Theta)$, used to describe the angular dependence of how light is scattered. Assumptions of phase functions vary widely in complexity \citep{hansen1969exact}. The most simplistic assumption is an isotropic scattering phase function. In this case, there is equal probability of arriving photons traveling to scatter in any given direction. Of course, scattering by gasses and particles is not isotropic. To account for more realistic phase functions, an asymmetry parameter, \emph{g}, is usually introduced, and used with more complex phase functions. It is important to note that asymmetry values vary widely depending on the specific optical properties of the condensing species \citep[e.g.,][]{morley2012neglected}. 

Figure \ref{fig:asy} shows distributions of asymmetry parameters and single scattering albedos from a wide range of giant planet models computed in \citet{batalha2018color} (data is available through \citet{batalha2018data}). All cloud models were computed using \citet{ackerman2001cloud} for a Jupiter-like system (1$\times$Solar metallicity, 25 ms$^{-2}$), with varying semi-major axes (i.e equilibrium temperatures). For reference, the asymmetry parameters of well-studied cirrus clouds and volcanic aerosols are also shown \citep{thomas2002radiative}. 

At 5 AU from a Sun-like star, H$_2$O/NH$_3$ clouds dominate the optical behavior, leading to high asymmetry values/single scattering albedo. At hotter temperatures more exotic cloud species such as ZnS and, Na$_2$S begin to widen and decrease the distribution of asymmetry values/single scattering albedos. Exoplanet atmospheres, which cover a broad range in mass and temperature space, will exhibit a wide range of optical properties. 

To emphasize how these wide ranges of optical properties propagate to the behavior of the phase function, Figure \ref{fig:hg} shows a typical phase function computed for a range in asymmetry parameters, $g$. 

The extreme range in these differences motivated the design of the new code. We aimed to create a code where fundamental radiative transfer assumptions, such as that of the phase function, could be easily assessed. We hope that this will guide development of future facilities, and facilitate a symbiotic relationship between future observations and the improvement of theoretical models of directly imaged exoplanets.

\subsection{Organization}
In what follows we describe the methodology of \texttt{PICASO} in \S\ref{sec:picaso}, with a special emphasis on the new physics/capabilities that have been introduced.  Then we will analyze the major assumptions made in our calculations of the reflected light in \S\ref{sec:datass} in order to show which assumptions the calculations are most sensitive to across a large parameter space in planet temperature, cloud composition, and stellar type. In \S\ref{sec:validate} we validate \texttt{PICASO} against two different independent calculations. Then, given the most up-to-date specifics of a future space-based direct imaging mission, we use \texttt{PICASO} to do a full information-content analysis across SNR and resolving power. Here, we will specifically focus on our ability to constrain atmospheric composition and gravity. We end with a discussion and conclusion in \S\ref{sec:discon}.

\section{\texttt{PICASO}: The Forward Model} \label{sec:picaso}
A full derivation of the radiative transfer of the forward model can be found \href{htpps://natashabatalha.github.io/picaso_dev}{online}.
\begin{figure}
\centering
\includegraphics[width=\columnwidth]{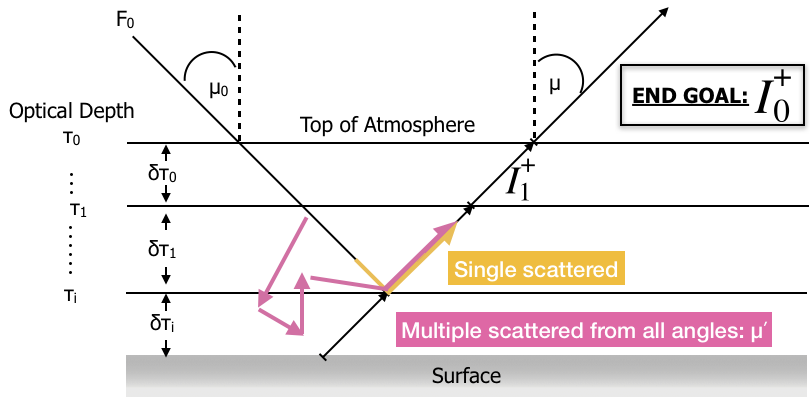}
\caption{Model schematic of \texttt{PICASO}. \href{https://natashabatalha.github.io/picaso_dev\#slide01}{\faSearch}}
\label{fig:scheme}
\end{figure}
As is with any atmospheric scattering code, we begin with the radiative transfer equation \citep{goody1989rte}: 

\begin{multline}
    I(\tau_i, \mu)= I(\tau_{i+1}, \mu)e^{\delta\tau_i/\mu} \\ - \int_0^{\delta\tau_i} S(\tau'\mu)e^{-\tau/\mu} d\tau`/\mu
\end{multline}

Here, the terms are as following: 
\begin{itemize}
    \item $I(\tau_i, \mu)$: the azimuthally averaged intensity emergent from the top of an atmospheric layer, $i$, with opacity, $\tau$, and outgoing angle, $\mu$
    \item $I(\tau_{i+1}, \mu)e^{\delta\tau_i/\mu}$: the incident intensity on the lower boundary of the layer attenuated by the optical depth \textit{within} the layer, $\delta\tau$
    \item $S(\tau',\mu)$: the source function, integrated over all layers
\end{itemize}
In our formalism, the source function only consists of two components: 1) the single-scattered radiation, and 2) the multiple scattered radiation, integrated over all diffuse angles. In other words, we do not include a thermal term in the source function. Traditionally, dating back to \citep{toon1989rapid}, the thermal and reflected light terms have been computed separately. We leave the addition of the thermal component to a future update so that the source function has the form: 
\begin{multline}
    S(\tau',\mu)=\frac{\omega}{4\pi} F_0P_{single}(\mu, -\mu_0)e^{-\tau'/\mu_s} + \\ \frac{\omega}{2} \int_{-1}^{1} I(\tau', \mu') P_{multi}(\mu, \mu') d\mu'
    \label{eqn:source}
\end{multline}
where the first term is the single-scattered radiation, whose behavior is described by the phase function $P_{single}(\mu, -\mu_0)$, and the second term is the multiple scattered radiation, whose behavior is described by $P_{multi}(\mu, \mu')$. A schematic of the plane-parallel model is shown in Figure \ref{fig:scheme}.

In addition to basic planetary properties (e.g. stellar spectrum, planet mass \& radius) \texttt{PICASO} takes in as input: 
1) a pressure-temperature profile and altitude-dependent abundances (see \href{https://natashabatalha.github.io/picaso/notebooks/1_GetStarted.html}{\faCode}, \texttt{justdoit.atmosphere()}), and 
2) a cloud profile (single scattering albedo, asymmetry parameter, and total extinction; see \href{https://natashabatalha.github.io/picaso/notebooks/2_AddingClouds.html}{\faCode}, \texttt{justdoit.clouds()}). As further shown in the tutorial, the cloud profile can either be input as a full altitude dependent profile parameterized or generated from a model such as \citet{ackerman2001cloud}, or it can be input as different cloud layers that are arbitrarily set by additionally supplying the cloud top pressures and vertical extent of each layer. \texttt{PICASO} is designed to accommodate several different input styles in order to be highly customizable for each user.

Our methodology is thoroughly described in \citet{cahoy2010exoplanet} (see Section 3.2). In short, we follow the source function method in \citet{toon1989rapid}. We first use the two-stream quadrature to solve for the diffuse scattered radiation. Then, we use the resulting two-stream intensity to approximate the source function. There are several other methods of solving this, e.g. $\delta$-M stream method \citep{wiscombe1977deltam}, which we will explore in a future release of the code. 

\texttt{PICASO} includes several ways of handling the single and multiple scattering phase functions, as compared with \citep{cahoy2010exoplanet}. Additionally, \texttt{PICASO} has been written to include a more physically motivated methodology for Raman Scattering. Therefore, we will devote \S\ref{sec:single}, \S\ref{sec:multi}, \& \S\ref{sec:raman} to these specific components and we reference \citet{cahoy2010exoplanet} for an explanation of the boundary condition formalism, which has not been altered. Finally, in \S\ref{sec:albedo} \& \S\ref{sec:geometry} we derive the methodology for computing various types of albedos and the planet phase geometry, respectively.

\subsection{The Single Scattering Component}\label{sec:single}

\begin{table*}[]
    \begin{center}
    \begin{tabular}{|c|c|c|c|} \hline
        \textbf{Key} & \textbf{Formalism} & \textbf{Inputs Required} & \textbf{Pro/Con} \\ \hline \hline
        \texttt{OTHG} & Eqn. \ref{eqn:OTHG} & $g$ & Does not capture back scattering.\\ \hline
        \texttt{TTHG} & Eqn. \ref{eqn:TTHG} & $g_f$, $g_b$, $c_1$,$c_2$,$c_3$ & Captures small back peak. \\ \hline
        \texttt{TTHG\_Ray}\footnote{\texttt{PICASO} default} & Eqn. \ref{eqn:TTHG_Ray} & $g_f$, $g_b$, $c_1$,$c_2$,$c_3$, $\tau_{ray}$,$\tau_{cld}$ & Captures sharper back scattering caused by Rayleigh. \\
        \hline
    \end{tabular} 
    \end{center}
    \caption{Single Scattering Options \href{https:/natashabatalha.github.io/picaso/notebooks/4_AnalyzingApproximations.html\#Direct-Scattering-Approximation}{\faCode}}
    \label{tab:single}
\end{table*}

For the direct/single scattering component, the most widely used form is the Henyey-Greenstein (HG) phase function because it is: a function of $g$, generally resembles ``real'' phase functions, and is non-negative for all values of $\Theta$. The one-term HG phase function has the form:
\begin{equation}
    p_{OTHG} = \frac{1-g^2}{(1 + g^2 -2g\cos\Theta)^{3/2}}
\label{eqn:OTHG}
\end{equation}
Here, $g$ is the asymmetry parameter, which is defined as: 
\begin{equation}
    g = \frac{1}{4 \pi} \int_{4\pi} p(\cos \Theta) cos \Theta d \omega
\end{equation} 
where $\Theta$ is the angle between the original direction and the scattered direction (related to the planet's phase, $\alpha$ function via $\alpha = \pi-\Theta$). By defining this parameter, we only need to determine the relative proportion of photons that are scattered in the forward versus backward direction (as opposed to each individual intensity). 

The asymmetry parameter, $g$ can be any value $-1 \le g \le 1$. In the limit when $g=1$, photons approximately continue traveling in their original direction, when $g=-1$ their directions are reversed, and when $g=0$ they are equally likely to travel in the forward or backward direction (i.e. isotropy, \href{https://natashabatalha.github.io/picaso_dev\#slide02}{\faSearch}).

Using Equation \ref{eqn:OTHG} for the single scattering phase function, is easily accessed in \texttt{PICASO} (\href{https:/natashabatalha.github.io/picaso/notebooks/4_AnalyzingApproximations.html\#Direct-Scattering-Approximation}{\faCode}), but it is not the default. This is because although Equation \ref{eqn:OTHG} captures the observed forward peak relatively well, it fails to capture the additional (albeit smaller) backward scattering peak that has been observed on the Moon, Mars, Venus and Jupiter \citep{sudarsky2005phase}. To account for this, a second term in the phase function can be introduced : 
\begin{multline}
        P_{TTHG}(\cos\Theta) = f P_{OTHG}(\cos\Theta,g_{f}) \\ + (1-f)P_{OTHG}(\cos\Theta,g_{b}).
\label{eqn:TTHG}
\end{multline}

Here, in addition to having two asymmetry factors ($g_f$ for the forward \& $g_b$ for the backward), we also have a new parameter, $f$, which describes the fraction of forward to back scattering. In \texttt{PICASO}, we  give $f$ the functional form of \begin{equation}
    f = c_1 + c_2g_b^{c_3}
\end{equation}
where users can specify $c_1$, $c_2$, and $c_3$. By default, \texttt{PICASO}, sets $g_f = \Bar{g}$, $g_b=-\Bar{g}/2$, and $f=1-g_b^2$. 
$\bar{g}$ is the cloud asymmetry factor that is computed directly from the cloud code \texttt{eddysed}, weighted by the contribution of cloud opacity ($\bar{g} = g_{cld} \tau_{cld}/\tau_{scat}) $\citep{ackerman2001cloud}. 

However these values are certainly not universal. Jupiter, Saturn, Uranus and Neptune all exhibit slightly different forward/back scattering peaks \citep{sudarsky2005phase,dyudina2016reflected}. For observations of exoplanets, these parameters will have to be fit for.  

The last component to consider is the effect of Rayleigh scattering, which acts to increase the back scattering peak. The Rayleigh phase function has the form: 
\begin{equation}
    P_{ray}(\cos\Theta) = \frac{3}{4} (1 + \cos^2\Theta) 
    \label{eqn:ray}
\end{equation}
In order to incorporate \textit{both} Rayleigh and the cloud scattering properties, we combine TTHG and Equation \ref{eqn:ray}, by weighting the two phase functions by the fractional opacity of each. In other words, if $\tau_{cld}$ is the contribution of scattering from clouds, $\tau_{ray}$ is the contribution of scattering from Rayleigh, and $\tau_{scat}$ is the total scattering, the phase function takes the form: 
\begin{equation}
    P_{TTHG\_ray} = \frac{\tau_{cld}}{\tau_{scat}}P_{TTHG} + \frac{\tau_{ray}}{\tau_{scat}}P_{Ray}
    \label{eqn:TTHG_Ray}
\end{equation}
This methodology was used in \citep{feng2018earth} and a similar methodology was employed in \citet{cahoy2010exoplanet}. It is also the default methodology used in \texttt{PICASO}. Table \ref{tab:single} has a full summary of the single scattering methodology. 

\subsection{Multiple Scattering Component}\label{sec:multi}

\begin{figure*}
\centering
\includegraphics[width=0.7\textwidth]{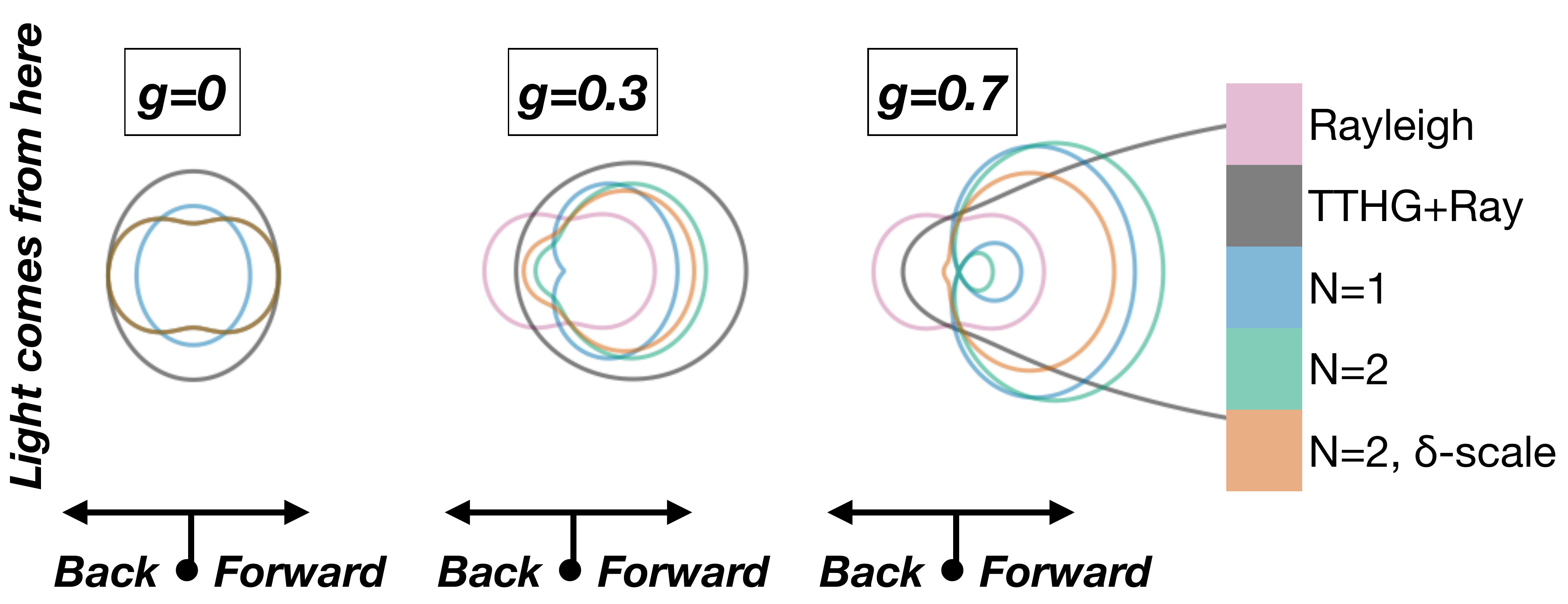}
\caption{All the options for multiple-scattering phase functions included within \texttt{PICASO}. TTHG with Rayleigh is shown for reference. \textbf{Main Point:} As asymmetry increases, approximations to the TTHG phase function get progressively worse. \href{https://natashabatalha.github.io/picaso_dev\#slide05}{\faSearch} \href{https://natashabatalha.github.io/picaso/notebooks/4_AnalyzingApproximations.html\#Multiple-Scattering-Approximations}{\faCode}}
\label{fig:multi}
\end{figure*}

We cannot use the same forms for phase functions, as we did in \S\ref{sec:single}, because the multiple scattering component of the source function (Equation \ref{eqn:source}) must be integrated over all diffuse angles, $\mu$. 

An additional convenience of the HG phase function, is that we can mathematically write it as a series of Legendre polynomials: 
\begin{equation}
    P_{multi}(\cos\Theta) \approx \sum_{l=0}^{N-1} \beta_lP_l(\cos\Theta)
\end{equation}
where $P_l(\cos\Theta)$ are the polynomials (not to be confused with another phase function), and $\beta_l$ are the moments of the phase function. The moments can be written out as:
\begin{equation}
    \beta_l = \frac{2l + 1}{2} \int_1^{-1} P_l(\cos\Theta)p(\cos\Theta)d\cos\Theta
\end{equation}
which should look familiar, given the previous equation shown for the asymmetry factor (see Equation 4). Therefore, the moment can just be simplified to $\beta_l = (2l + 1)g_l$. 

Expanding this polynomial to simply an $N=1$ expansion gives us: 
\begin{multline}
    P_{multi}(\cos\Theta) = 1 + 3 \bar{g}\cos\Theta = 1 + 3  \bar{g} \mu\mu'
    \label{eqn:n1}
\end{multline}
where an azimuthal independence assumption can reduce $\cos\Theta=\mu\mu'$. 

One downfall to this formalism (similar to the OTHG) is that it fails to capture the behavior of Rayleigh scattering (as $\bar{g}$ is zero for Rayleigh). Therefore, many authors \citep[originating from][]{snook1999optical}, have leveraged the fact that the second order Legendre dependence on $\cos^2\Theta$ is the same as that of Rayleigh (see Equation \ref{eqn:ray}). Therefore, by forcing the second moment $\beta_l=(2l + 1)g_l=g_2$ to yield the Rayleigh phase function, we can accurately account for Rayleigh scattering. We set this new parameter, $g_2=\tau_{ray}/(2\tau_{scat})$, so that when Rayleigh dominates the total opacity, $g_2$ approaches 1/2, the correct value for Rayleigh \citep{hansen1974light}.

For multiple scattering, the two options for \texttt{PICASO} are \texttt{N=1} and \texttt{N=2} expansions (\texttt{N=2}, being the default). However, we can also add $\delta$-Eddington methodology (explained in the following section) to further improve accuracy.

\subsection{$\delta$-Eddington}
Low order Legendre expansions are not adequate enough to represent very high forward scattering (which is very asymmetric). Given the high scattering asymmetry produced by Mie scattering particles with sizes larger than typical optical wavelengths  (Figure \ref{fig:asy}), this could be problematic. In order to make lower order approximations more accurate, \texttt{PICASO} leverages the $\delta$-Eddington Approximation \citep{joseph1976deltaedd}. In this approximation, $\bar{g}, \tau, \omega$ (the single scattering albedo) are all scaled by recognizing that a beam which experiences a high degree of forward scattering from high albedo particles essentially still propagates forward with little alteration:
\begin{multline}
    g'=\frac{\bar{g}}{1+\bar{g}}, \tau' = \tau(1-\omega\bar{g}^2), \omega' = \frac{\omega(1-\bar{g}^2)}{1-\omega\bar{g}^2}
\end{multline}
Figure \ref{fig:multi} shows a full comparison of all the multiple scattering phase functions. Note, the $N=2$ expansion with $\delta$-scaling reduces the forward peak to regions where the Legendre polynomials should be in higher agreement with the TTHG (i.e. lower asymmetry). 

\subsection{Raman Scattering}\label{sec:raman}
A small fraction of photons that are scattered via the well-known Rayleigh process, experience a shift to redder wavelengths caused by the excitation of rotational and vibrations transitions in atmospheric gasses. Although some incident stellar photons are shifted, when we compute the albedo, we normalize by the original incident flux (see Equation \ref{eqn:albdo}). This discrepancy between the shifted incident radiation and the original incident radiation creates new spectral features in the albedo calculation. These detectable shifts are called ``ghost'' features in the reflected light spectra of planetary atmospheres \citep{price1977raman}. 

Raman scattering has been detected in the reflected light of all Solar System gas giants \citep[e.g.][]{karkoschka1994spec, yelle1987analysis,courtin1999raman}. Recently, it was also suggested that Raman scattering could be an important indicator of the main spectroscopic scatterer in the atmospheres of exoplanets, such as H$_2$ vs. N$_2$ \citep{antonija2016raman}. Additionally, it was shown that varying stellar spectra will have a non-negligible affect on the reflected light of exoplanets \citep{antonija2017stellar}. 

For the studies of exoplanets, the effect of Raman scattering has been approximated using the methodology of \citet{pollack1986estimates} \citep[e.g.][]{marley1999reflected, sudarsky2005phase, cahoy2010exoplanet}. All of these analyses computed Raman scattering correction terms for a 6000~K blackbody. The \citet{pollack1986estimates} approximation captures the overall shape of Raman scattering by H$_2$ (i.e. decreased reflectively toward the blue). However, it fails to capture specific `ghost' features from the stellar spectrum at higher resolving powers (as the shift of individual stellar spectral lines are resolved). 

For \texttt{PICASO}, we modify the \citet{pollack1986estimates} approximation to include the Raman cross sections of H$_2$ computed by \citet{antonija2016raman}. We also retain the original \citet{pollack1986estimates} methodology, as an option, for low resolution, low SNR observations. 

Following \citet{pollack1986estimates} we introduce the effect of Raman scattering by adding a correction term to the Rayleigh opacity, $\tau_{\rm Ray}$.
\begin{equation}
   f_{\rm Ram} = \frac{\sigma_{\rm Ray} + \sigma_{\rm Ram}(f_{\lambda*}/f_{\lambda})}{\sigma_{\rm Ray} + \sigma_{\rm Ram}} 
   \label{eqn:raman}
\end{equation}
where $\sigma_{\rm Ram,Ray}$ are the cross sections of both Raman and Rayleigh scattering, respectively, and $f_\lambda,\lambda*$ are the solar spectra at unshifted, and shifted wavelengths. Each excitation corresponds to a specific wavelength shift of $\lambda^{*-1} = \lambda^{-1} + \Delta \lambda^{-1}$, where $\Delta \lambda$ is the wavelength shift. For reference, the strongest transition of H$_2$ (the vibrational fundamental) is $\Delta \nu=$ 4161 cm$^{-1}$.  

Unlike \citet{pollack1986estimates}, we use stellar spectral models from \citet{castelli2004grid} for this analysis. \texttt{PICASO} uses \texttt{PySynphot} \citep{pysynphot2013} so that users can draw from different stellar databases.

Additionally, we include initial rotational levels ranging from J=0 to J=9 for H$_2$ only since those were the transitions provided by \citet{antonija2016raman}. The cross sections for any given transition from an initial quantum state of $v=0, J_i$, to final quantum state of $v_f, J_f$ is given by (see Equation A4; \citet{antonija2016raman}): 
\begin{equation}
    \sigma_{\rm Ram}(0,J_i,f_f,J_f,\lambda) = \frac{C}{\lambda*\lambda^3} \text{[cm$^2$]}
\end{equation}
The constant $C$ is given by the values in Table A1 of \citet{antonija2016raman}. 

In a future update we will include Raman scattering by N$_2$ and He but for this work, H$_2$ is sufficient to study the approximate behavior or Raman scattering.

\subsection{Computing Different Types of Albedos}\label{sec:albedo}
\texttt{PICASO} computes three different kinds of albedos: spherical albedo ($A_s$), geometric albedo ($A_g$), and the Bond albedo ($A_b$). The spherical albedo, $A_s$, is the fraction of incident light reflected by a sphere towards all angles, and it can be computed for a planet phase geometry, $\alpha$, by: 
\begin{equation}
    A_s(\lambda) = 2 \int^\pi_0 \frac{F_p(\alpha,\lambda)}{F_{0,L}(\lambda)} \sin \alpha d\alpha \label{eqn:albdo}
\end{equation}
where $F_p$ is the emergent flux from the planet, and $F_{0,L}$ is the flux from a perfect Lambert disk under the same incident flux, $F_0$. This spherical albedo, integrated over all angles can be written in two parts: the geometric albedo, and the phase integral. The geometric albedo is just the ratio of the reflected planet flux at full phase, to the incident flux from the perfect Lambert disk:
\begin{equation}
     A_g(\lambda) = \frac{F_p(\alpha=0^\circ,\lambda)}{F_{0,L}(\lambda)}.
\end{equation}
Then, the phase integral, which is normalized to be 1.0 at full phase, can be written as:
\begin{equation}
    q = 2 \int^\pi_0 \frac{F_p(\alpha,\lambda)}{F_p(\alpha=0^\circ,\lambda)} \sin \alpha d\alpha.
\end{equation}

Although we do not show any Bond albedos here, \texttt{PICASO} does contain the functionality to compute it. The Bond albedo is a stellar flux-weighted reflectivity that is integrated by wavelength: 
\begin{equation}
    A_b = \frac{\int_0^\infty A_s(\lambda) F_0(\lambda)  d\lambda}{\int_0^\infty  F_0(\lambda) d\lambda}.
\end{equation}
Therefore, Bond albedos will vary for two planets that have identical spherical/geometric albedos but orbit different stars \citep{marley1999reflected}. Since a directly imaged exoplanet will never be observed at full phase, the traditional geometric albedo (which arose from Solar System heritage) can be somewhat cumbersome, given the ratio to the
ideal Lambert disk. Nevertheless for ease of comparison with the existing literature we here report results primarily in this framework.

\subsection{Planet Phase Geometry}\label{sec:geometry}
In order to capture phase-dependence, we compute the emergent intensity from the disk at several plane-parallel facets, where each facet has its own incident and outgoing angles. Following \citet{horak1965calculation}, we use a Chebyshev-Gauss integration method to integrate over all the emergent intensities \citep[also used in][]{cahoy2010exoplanet, madhu2012analytic,webber2015effect}. By default, \texttt{PICASO} includes 10 Chebyshev and 10 Gauss angles, which strikes a balance between computational speed and physical accuracy. However this can easily be modified in the code (see options in \texttt{justdoit.phase\_angle()}). Of course, increasing the number of angles, increases compute time. However, if the user is particularly interested in capturing scattering at high cosine angles (e.g. near the limbs), then it is necessary to increase the number of integration angles accordingly.

Chebyshev-Gauss angles easily translate to planetary latitude and longitude, making it possible to explore the effect of 3D general circulation models on albedo spectra, as in e.g. \citet{webber2015effect} and \citet{lee2017dynamic}. Although we do not currently include this in the set of \texttt{PICASO} tutorials, we will make this \texttt{jupyter notebook} available soon.

\begin{figure*}
\centering
\includegraphics[width=0.9\textwidth]{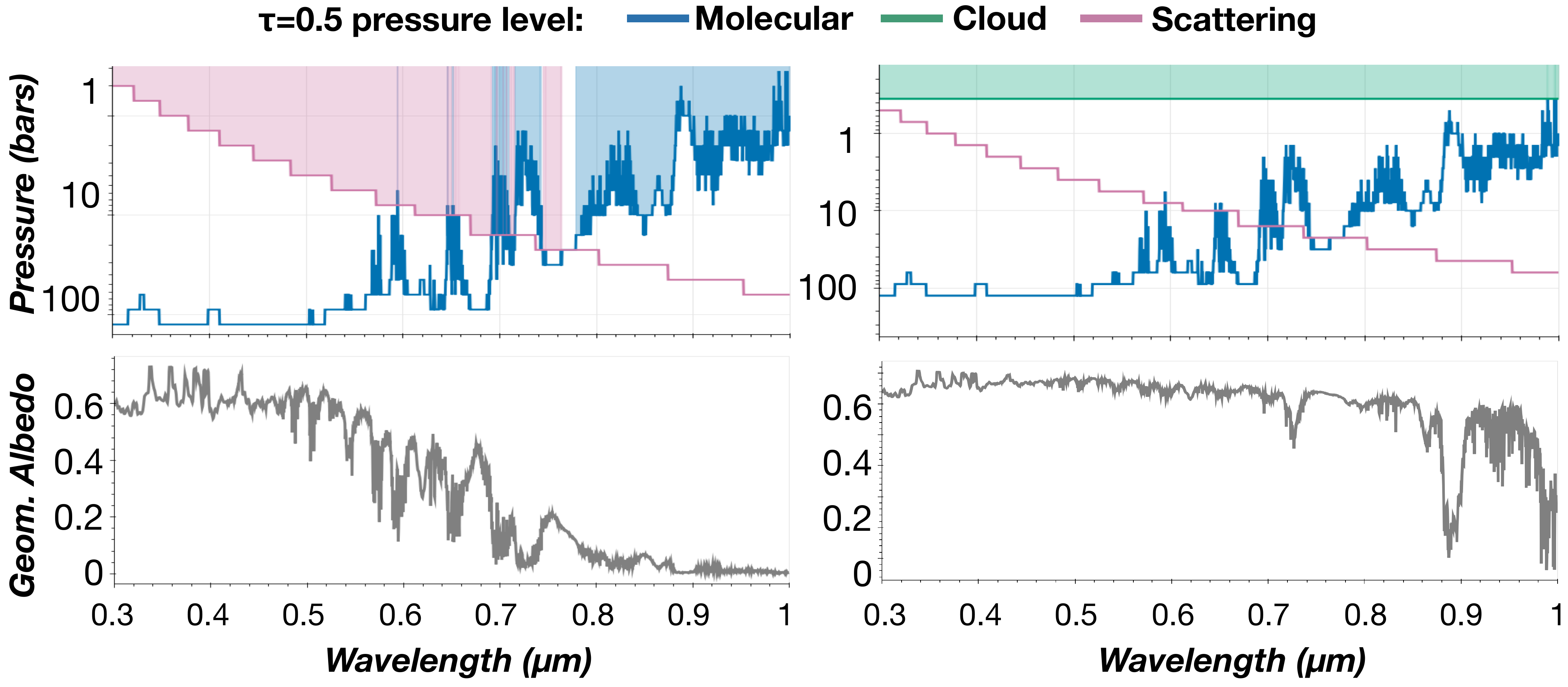}
\caption{\textit{Photon attenuation plot}, which shows the pressure level at which the two way optical depth encountered by a photon traversing the atmosphere at $\mu=0.5$ is $\tau=1$ (top panel). The lower panel shows the corresponding albedo spectra. Each spectrum is computed for a $25\,\rm m\,s^{-2}$ Jupiter-like planet, 5~AU from a Sun-like star. The left panel is a cloud-free model, while the right panel contains a cloud profile modeled with a sedimentation efficiency of $f_{\rm sed}=3$.  \textbf{Main Point:} Photon attenuation plot shows the dominant source of reflectivity, and therefore can give insight into the overall shape of the reflected light spectrum.}
\label{fig:atten}
\end{figure*}

\section{An Analysis of the Modeling Assumptions}\label{sec:datass}

Given our reflected light model, we now aim to determine which modeling assumptions are most important across a large range in parameter space. In particular, we are interested in sampling a parameter space across approximate temperature, cloud properties, and stellar spectrum (to test effects of Raman scattering). 

In order to do so, we require, as input, the temperature-pressure profiles, cloud structure, atmospheric composition profile. \citet{batalha2018color} created a large grid across this particular parameter space. It covered, for a planet with a gravity of 25 m/s$^2$, planets with semi-major axes ranging from 0.5-5 AU around a Sun-like star, metallicities (M/H) of 1-$100\times$Solar, and cloud profiles ranging from $f_{\rm sed}=0.01-6$. 

We use these models published in \citet{batalha2018color} as input. Briefly, the temperature profiles were computed using the radiative-convective model initially developed by \citet{mckay1989thermal} and later updated by \citet{marley1999thermal,marley2002clouds,fortney2005comp, fortney2008unified}. 

The cloud profiles were computed using a Mie scattering treatment of particle sizes calculated from the model developed by \citet{ackerman2001cloud}. Each profile was computed using a specific value of $f_{\rm sed}$, which is used to tune the sedimentation efficiency of the atmosphere. High values of $f_{\rm sed}>1$, produce vertically thin clouds with large particles, low values of $f_{\rm sed}<1$, produce the opposite--vertically thick clouds with small particles. The \citet{ackerman2001cloud} produces, as output: single scattering albedo, cloud extinction, and asymmetry values as a function of atmospheric layer, and wavelength. 

The top panel of Figure \ref{fig:atten} contains two plots which show the depth in the atmosphere at which the two way optical depth encountered by a photon traversing the atmosphere at $\mu$=0.5 is $\tau=1$, which we henceforth denote as a \textit{photon attenuation plot}.  These two models, chosen from our grid, are computed at 1$\times$Solar with a semi-major axis of 5~AU. The left panel contains a model of a cloud-free system, while the right contains a system with a cloud sedimentation efficiency of f$_{\rm sed}$=3. We showcase these two case studies throughout the analysis of the modeling assumptions.

The shaded regions of the photon attenuation plot indicate what the dominant source of opacity is as a function of wavelength: molecular absorption (blue), cloud absorption and scattering (green), or Rayleigh scattering (pink). Over the full illuminated hemisphere of a planet the angle of incidence of course varies from $\mu=0$ to 1 and scattering within cloud decks can increase the effective path length of a photon through the absorbing gas. Thus no single plot can fully capture the complete complexity inherent in the problem, but we find plots such as these helpful for understanding how the shape of reflected light spectra can be traced back to the dominant sources of reflectivity and absorption in the atmosphere. This is especially true for cases where you may have an interplay between the muting of strong scattering features (e.g. Raman) from the presence of optical absorbers (e.g. Na \& K).

Indeed such plots are commonly used in Solar System planetary science to help illuminate the relative importance of scattering and absorption at different wavelengths \citep[e.g.,][]{sromovsky2009uranus}.

\subsection{Sensitivity to Single Scattering Phase Function}
\begin{figure*}
\centering
\includegraphics[width=0.9\textwidth]{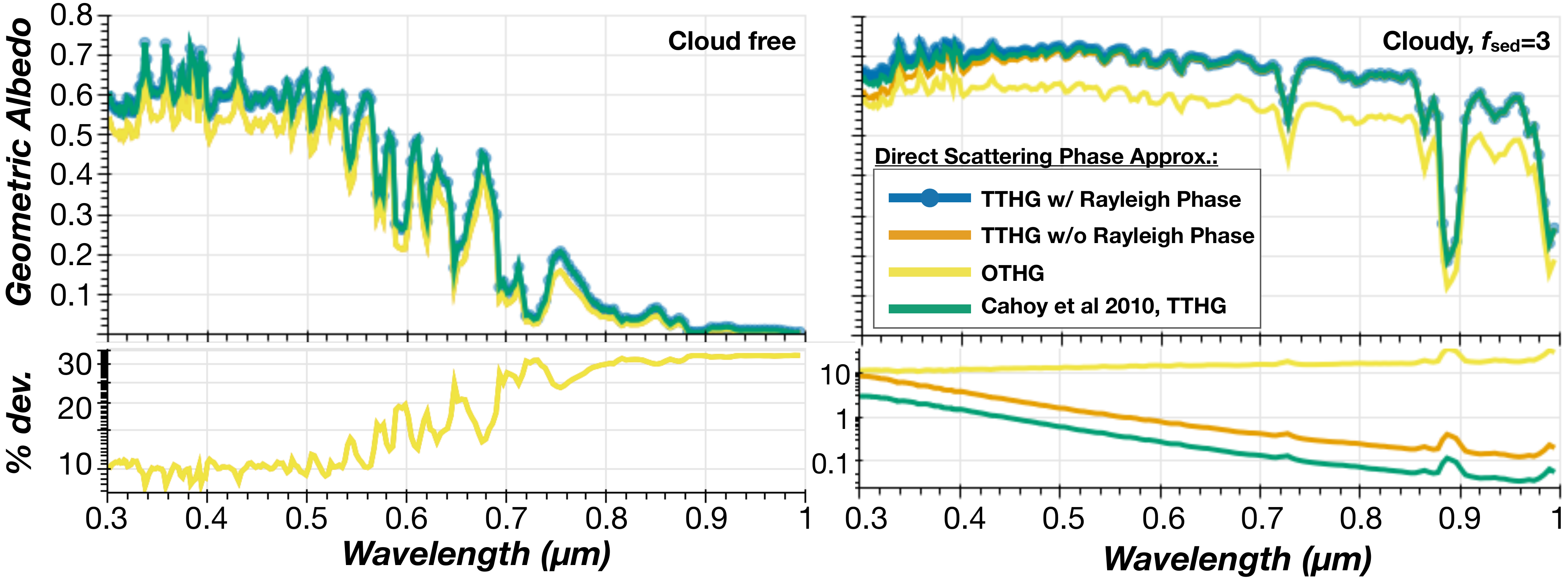}
\caption{\texttt{PICASO}'s four different methodologies for direct scattering phase functions. All spectra are computed at full phase for a 25m/s$^2$ Jupiter-like planet 5~Au from a Sun-like star. The left panel is a cloud-free model, while the right panel contains a cloud profile modeled with f$_{sed}$=3. \textbf{Main Point:} Assumptions of direct scattering phase function have large effect ($\le$30\%) on the resultant spectrum across all wavelengths. \href{https://natashabatalha.github.io/picaso_dev\#slide02}{\faSearch} \href{https://natashabatalha.github.io/picaso/notebooks/4_AnalyzingApproximations.html\#Direct-Scattering-Approximation}{\faCode}}
\label{fig:singlespec}
\end{figure*}
We first explore the sensitivity of \texttt{PICASO} to the choice in single scattering phase function. Figure \ref{fig:singlespec} shows the same planet case as that in Figure \ref{fig:atten}. The same spectrum was run using each of the four ways of representing direct scattering in \texttt{PICASO}. 

For a cloud-free case where there are no highly asymmetry scatterers, the two TTHG functions and the function used in \citet{cahoy2010exoplanet} are all identical (since $g_{cld}=0$). However, the cloud-free case shows deviations from the code default (\texttt{TTHG\_ray}) on the order of 10-30\% when the \texttt{OTHG} phase function is used.

For the cloudy case, Cahoy et al.'s phase function closely matches \texttt{TTHG\_ray}, except when Rayleigh scattering opacity is high toward the blue where deviations of $\le10$\% are present. Note all deviations are strongly sensitive to wavelength. 

\texttt{OTHG} exhibits the greatest deviation from the other phase functions, because it is not accounting for the back scattering peak from Rayleigh scattering. Figure \ref{fig:hg} shows that the Rayleigh contribution as a small back scattering contribution. It is important to note that the actual Rayleigh phase function is symmetric forward and back. But when you combine it with forward scattering particles, the net scattering is more forward.

Even for cases that are seemingly less asymmetric (semi-major axis, $a_s$=0.5 AU, see Figure \ref{fig:asy}), the specification for direct scattering phase function can still produce spectra that have maximum differences on the order of 100\% for full phase observations, and 50\% for phase=90$^\circ$. 

Although we do not show the specific effect of changing, $f$, the fraction of forward to back scattering, it will also strongly impact the resultant spectrum. Smaller fractions will produce smaller back scattering peaks and yield significantly dimmer spectra across wavelength, and vice versa. 

\noindent \textbf{Modeling Recommendation:}
\begin{itemize}
    \item Use default specification for direct scattering (\texttt{TTHG\_Ray}) 
    \item Fit for the functional form of the fraction, $f$, of forward to back scattering according to the problem being addressed. 
\end{itemize}

\subsection{Multiple Scattering Phase Function \& $\delta$-Eddington}
\begin{figure*}
\centering
\includegraphics[width=0.9\textwidth]{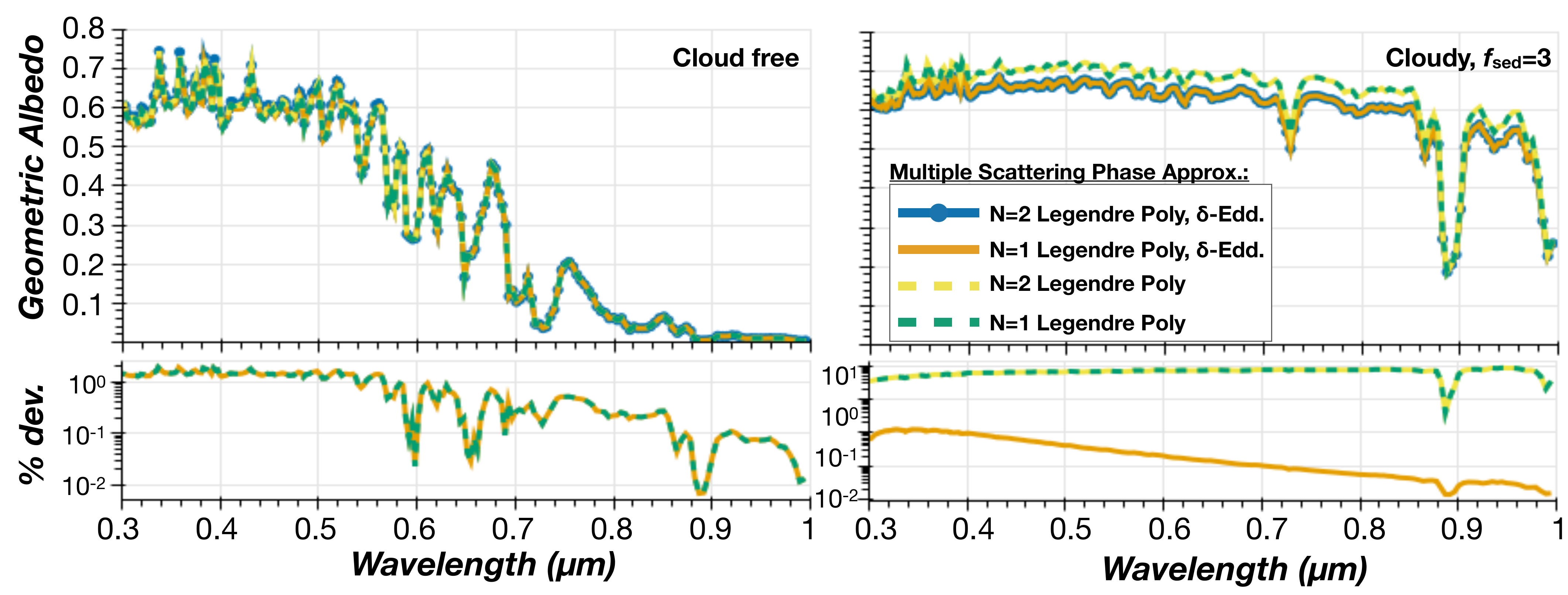}
\caption{Spectra and percent deviation between the code default and all other \texttt{PICASO} specifications for phase function. Planet cases are identical to those shown in Figure \ref{fig:singlespec}. \textbf{Main Point:} Use of $\delta$-Eddington approximation has largest impact on spectra.\href{https://natashabatalha.github.io/picaso_dev\#slide04}{\faSearch} \href{https://natashabatalha.github.io/picaso/notebooks/4_AnalyzingApproximations.html\#Multiple-Scattering-Approximation}{\faCode}}
\label{fig:multispec}
\end{figure*}
Next, we assess \texttt{PICASO}'s sensitivity to the multiple scattering phase function, and the $\delta$-Eddington approximation. Figure \ref{fig:multispec} shows the modeling sensitivity to the user's choice for multiple scattering phase function. For both cloud-free and cloudy cases, there are $\le1$\% differences when choosing between a $N=1$ or $N=2$ Legendre expansion. The $N=2$ expansion is used to approximate the multiple scattering by Rayleigh scattering. Therefore, for the cases modeled here, the diffuse scattering by Rayleigh is a relatively small contribution to the total reflectivity. As observations increase in precision we will have to revisit whether or not this holds true. Studies of the accuracy of Legendre polynomial expansions suggest that they made degrade in accuracy for asymmetric large particles \citep{zhang2017comparison}. \citet{zhang2017comparison} also suggested that Chebyshev polynomial expansions as a more accurate alternative. We will save this for a future update, when better data warrant higher accuracy phase functions. 

In order to improve the parameterization of these expansions, \texttt{PICASO} leverages the $\delta$-Eddington method of scaling the single scattering albedo, opacity, and asymmetry parameter. When $g_{cld}$ is nonzero, choice of the $\delta$-Eddington method impacts the spectra by up to 30\% in some cases. We set $N=2$ $\delta$-Eddington as default since it can reproduce, with relatively high accuracy, observations of Earth \citep{feng2018earth}, and Jupiter \citep{cahoy2010exoplanet}.

As more diverse populations of exoplanets are observed in reflected light with higher signal-to-noise ratio (SNR), we will conduct a more thorough investigation of these approximations. Since the publication of the $\delta$-Eddington method \citep{joseph1976deltaedd}, several other techniques have also been developed to improve the phase function parameterization \citep[e.g.][]{hu2000delta,iwabuchi2009fast,sorensen2017q}. We will consider these in a future update. 

\noindent \textbf{Modeling Recommendation}: For planet cases with some degree of asymmetric cloud scatterers, always use N=2 Legendre polynomial expansion with the $\delta$-Eddington correction. 
\subsection{Raman Scattering}
\begin{figure*}
\centering
\includegraphics[width=0.9\textwidth]{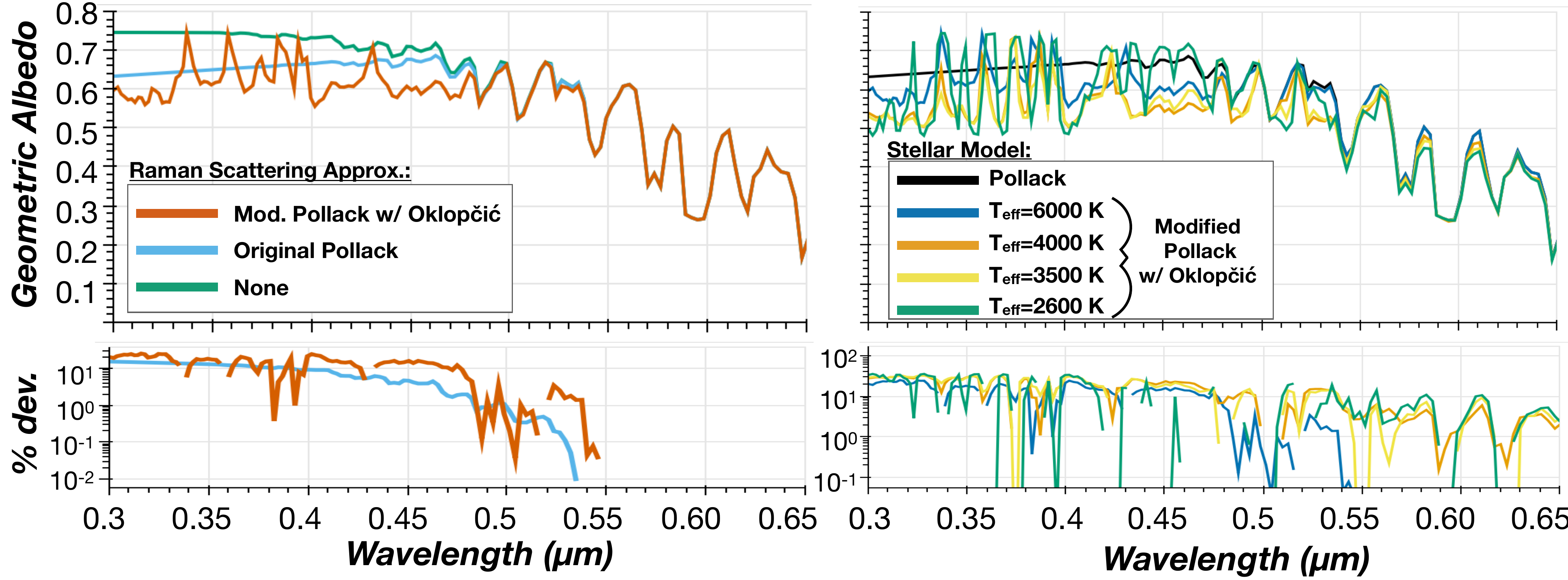}
\caption{Spectra and percent deviation between the code default and all other \texttt{PICASO} specifications for phase function. Planet cases are identical to those shown in Figure \ref{fig:singlespec}. \textbf{Main Point:} Even at low resolution (R=100), Raman scattering is important to include, especially for cool stars.\href{https://natashabatalha.github.io/picaso/notebooks/4_AnalyzingApproximations.html\#Raman-Scattering-Approximations}{\faCode}}
\label{fig:ramanspec}
\end{figure*}
Figure \ref{fig:ramanspec} shows the modeling sensitivity to \texttt{PICASO}'s two methodologies for computing Raman scattering. The \citet{pollack1986estimates} approximation captures the general behavior of the decline in reflectively toward the blue, but fails to produce any spectral features. When the Pollack et al. approximation is modified to include cross sections computed from \citet{antonija2016raman}, ghost spectral features are introduced at the $\sim10$\% level. Spectral features begin to disappear at R$\sim$50, but small 1\% baseline differences still remain at R$\sim$10.

For a $T_{\rm eff}$=6000~K star (the Pollack et al. default), 10\% differences remain through 0.55$\mu$m (after the Mg I feature at 5200$\AA$). Cooler stars (toward $T_{\rm eff}$=2600~K), with more crowded molecular features, create spectral differences out past 0.65$\,\mu$m. Cloudy spectra (e.g. Figure \ref{fig:multi} right panel) are also sensitive to Raman scattering features despite the prominent cloud opacity in the blue. 

Differences between the calculations here and those shown in \citet{antonija2017stellar} (see Figure 4) and \citet{sromovsky2005accurate} (see Figure 17) are attributed to the resolution of the stellar spectrum, and the stellar databases chosen. A key input to modeling Raman scattering correctly is an accurate, high-resolution stellar spectrum. \citet{antonija2017stellar} used stellar spectra from \citet{valdes2004stellar} database, which are computed with $\Delta\lambda$=1~$\AA$. \citet{sromovsky2005accurate} used a Solar spectrum from the Upper Atmospheric Research Satellite, which had a nominal resolution of 2~$\AA$.
\begin{figure*}
\centering
\includegraphics[width=0.9\textwidth]{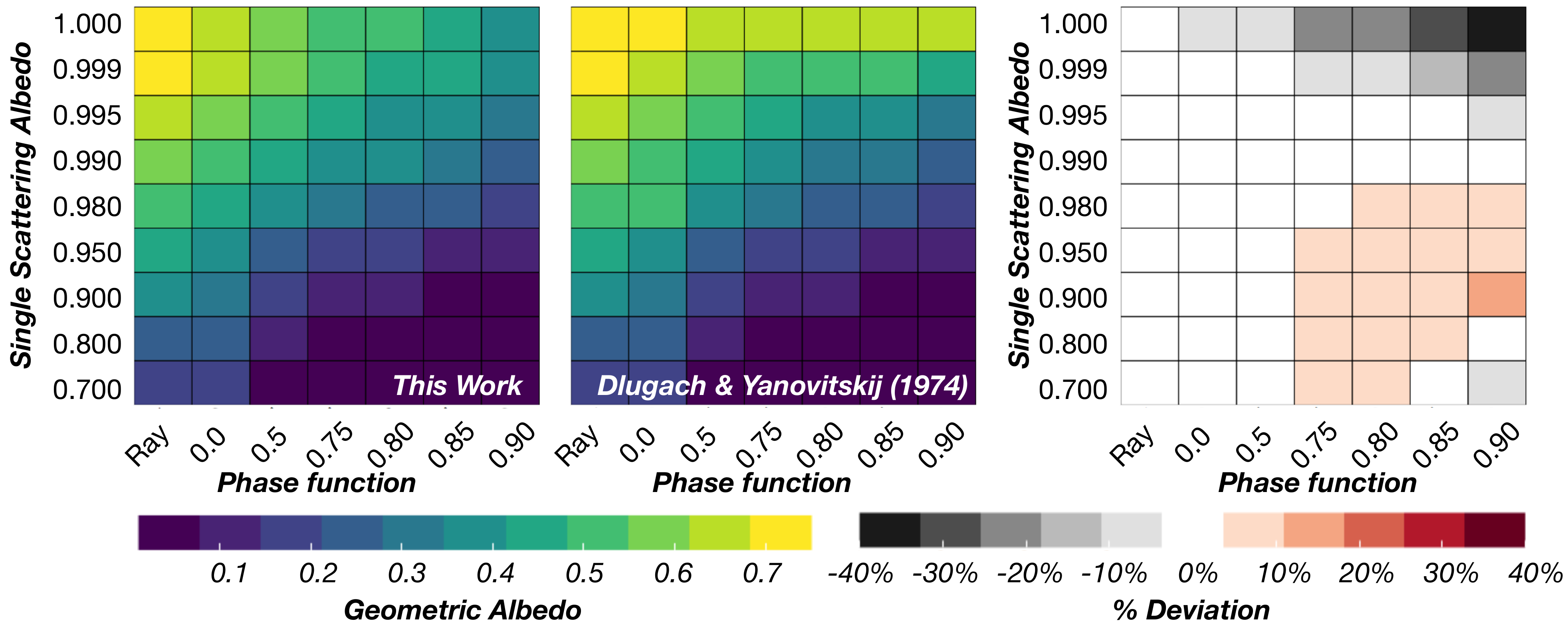}
\caption{A summary of the benchmark analysis between \citet{dlugach1974optical} and this work. Left two panels shows the geometric albedo for a range of single scattering albedos and phase functions. Right-most plot shows the difference map between the two. \textbf{Main Point}: Models agree within 10\% for most cases, with a maximum discrepancy of 40\% for highly asymmetric cases.}
\label{fig:base1}
\end{figure*}
\begin{figure}
\centering
\includegraphics[width=0.9\columnwidth]{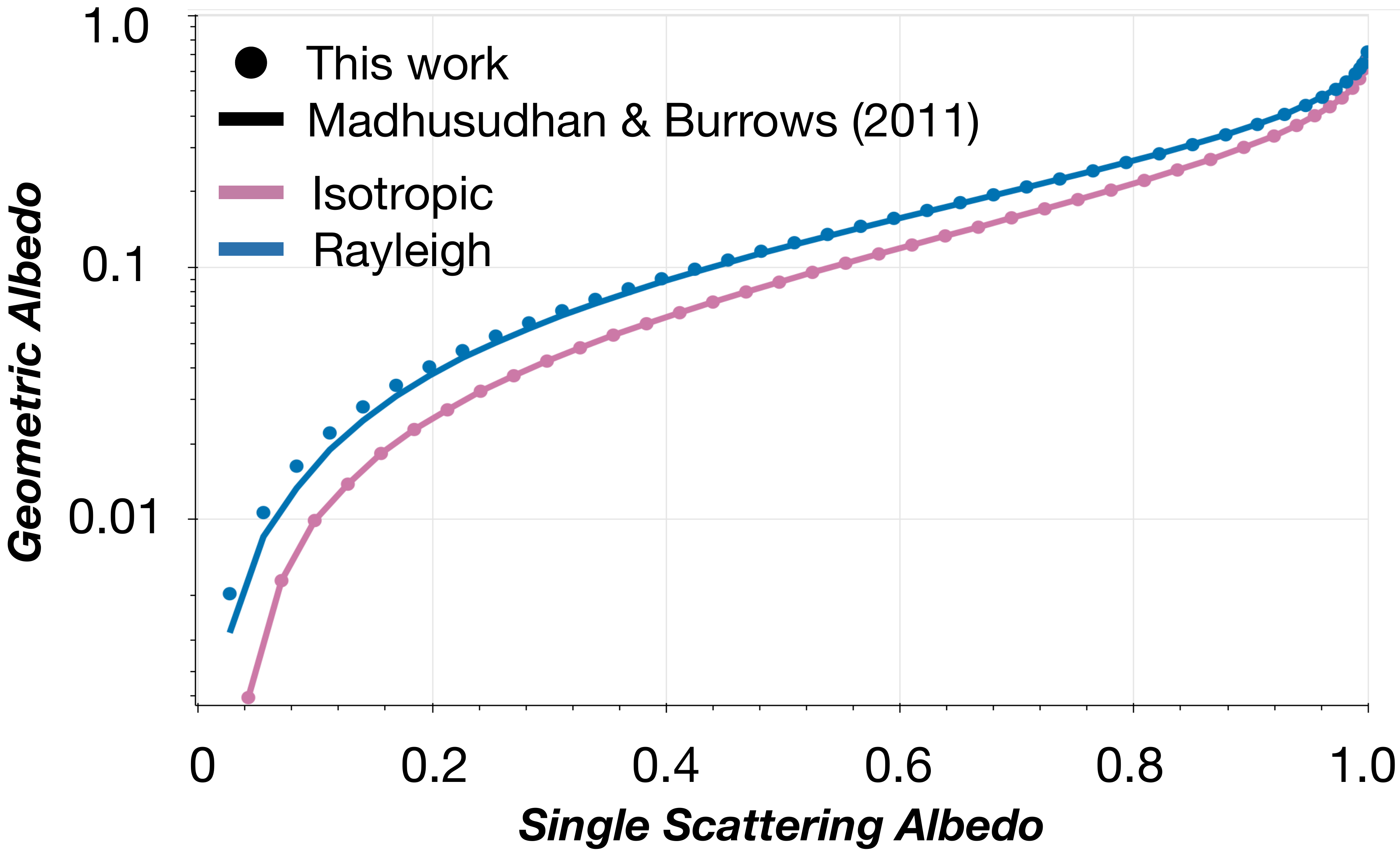} 
\caption{Summary of the benchmark analysis between \citet{madhu2012analytic} and this work. The greatest divergence occurs at very low geometric albedo ($\le$0.1). \textbf{Main Point:} Overall, models agree within 10\% in all isotropic and rayleigh phase functions cases.}
\label{fig:base2}
\end{figure}
The \citet{castelli2004grid} grid used here is computed at a resolution of 10~$\AA$. This lower resolution grid will result in an under estimation of the Raman effect. One additional, minor, difference can be seen in the $T_{eff}$=2600~K spectrum. Around 0.35~$\mu$m, some spectral features appear to have flat tops. This is a result, originally pointed out in \citet{courtin1999raman}, of instabilities in the solution of the radiative transfer equation that prevent us from allowing Equation \ref{eqn:raman} to be greater than 1. Despite these differences, our modified Pollack approximation is a much more accurate solution than the original Pollack methodology.

One last subtlety is that Figure \ref{fig:ramanspec} makes it seem as if Raman scattering has a dramatic effect on the total energy of budget of the atmosphere. This is somewhat exaggerated by the way in which the albedo is defined, by ratioing to the stellar flux. It has to be accounted for in order for the ratio to be correct, but in actuality is not a huge influence on the energy budget of the atmosphere. The effect of the small deposition of energy into the atmosphere by the small wavelength shifts that occur for those photons that experience this form of scattering would have to be computed by a complete radiative-convective equilibrium code which carefully tracks the energy budget of the atmosphere. Since \texttt{PICASO} is focused on tracking the reflectivity of the atmosphere as a whole it is not well suited to this particular task and doing so is beyond the scope of this paper.

\noindent \textbf{Modeling Recommendation}: Choose Pollack et al. methodology with \citet{antonija2016raman} cross sections, and choose a stellar spectrum that matches the level of resolution and accuracy needed. 

\section{Benchmark Analysis}\label{sec:validate}
In order to benchmark the accuracy of the code, we chose to compare against the results of \citet{dlugach1974optical} and \citet{madhu2012analytic}. \citet{dlugach1974optical} computed the intensity of radiation diffusely reflected from a semi-infinite homogeneous atmosphere with arbitrary single scattering phase function. Their analysis focused on the optical properties of Venus and the Jovian planets. Therefore, they carried out calculations for Rayleigh and the HG phase functions with asymmetry parameters ranging from 0-0.9, and single scattering albedos ranging from 0.7-1. \citet{madhu2012analytic} provided analytic phase expressions for geometric albedo as a function of single scattering albedo for both Rayleigh scattering and isotropy in a semi-infinite atmosphere. We compare \texttt{PICASO} against two models (one originating from Solar System science, the other originating from exoplanet science) across a wide range in phase function, and single scattering albedo is sufficient enough to prove accuracy of the model. 

Figure \ref{fig:base1} shows the first comparison against \citet{dlugach1974optical}. Dlugach et al. used a one-term HG phase function for all asymmetric calculations. Therefore, it is important to note that if comparisons are carried out using \texttt{PICASO}'s default (as opposed to using \texttt{OTHG} for the single scattering phase function), the results will not agree well. This further motivates our choice for inheriting older methodologies of computing phase functions so that fruitful code comparisons are easily accessible. 

Using a \texttt{OTHG} phase function, the models agree within 10\% for all Rayleigh phase functions and for $g\le$0.5. The models start to exhibit 10\% differences for $0.5< g \le$0.85. Since the diversity of cases illustrated in Figure \ref{fig:asy} fall in this range of asymmetry values, we feel that this can be considered in good agreement. There are few cases with $g=0.9$ that exhibit $\sim$40\% differences. However, it is not obvious what these differences could be attributed to. \citet{dlugach1974optical} computes higher geometric albedos when single scattering is less than 0.98, and lower geometric albedos when single scattering is $\sim$1. Given the complete independence of the two models, there are numerous factors that could potentially contribute to this including: the diffuse scattering calculation, geometric integration, radiative transfer solver. Because majority of cases fall within 10\% agreement we consider these two models to be in good agreement. 

Figure \ref{fig:base2} shows the comparison between the calculations in  \citet{madhu2012analytic} and \texttt{PICASO}. Here, we only compare isotropic and Rayleigh cases across a range of single scattering albedos. Our results are well within 10\% agreement. The largest deviation comes from the computation of very low geometric albedos ($\sim$0.01). Such very low single scattering albedos are well outside the range expected for the types of clouds expected (Figure \ref{fig:asy}), although unusual composition particles (e.g., \citet{gao2017sulfur}) can be quite dark at some wavelengths. We consider \texttt{PICASO} and the analytic model of \citet{madhu2012analytic} to be in good agreement.  

\section{Information Content Analysis of Reflected Light}\label{sec:ic}
\begin{figure*}
\centering
\includegraphics[width=0.8\textwidth]{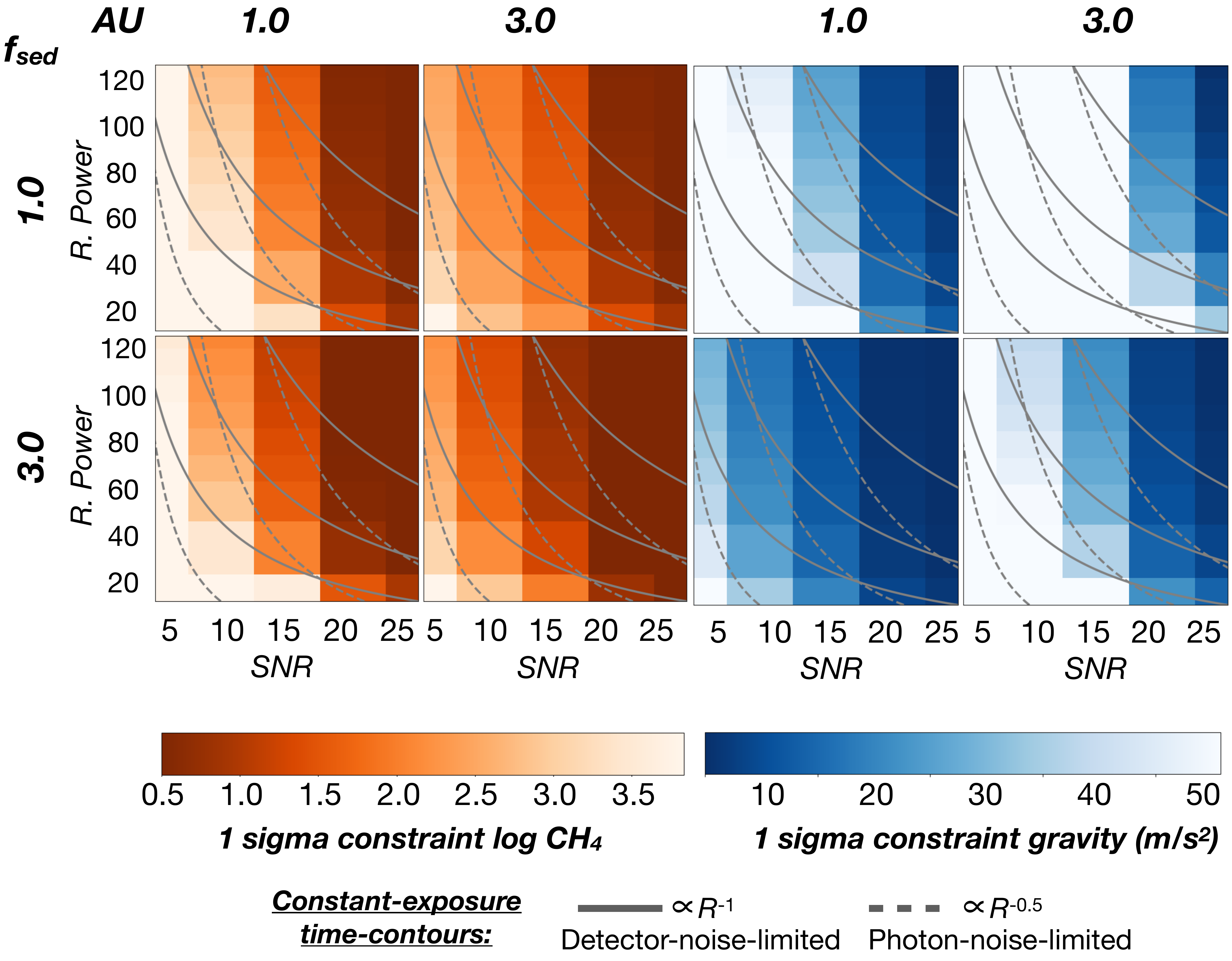} 
\caption{Information content analysis for a subset of the full parameter space covered here. In all plots, darker colors correspond to tighter constrained planetary systems. All spectra simulations were computed for a 1$\times$Solar, 25m/s$^2$ planet, around a Sun-like star. The approximate \textit{WFIRST} wavelength range of 0.5-0.76$\mu$m was used with the SNR calculations computed in \citet{nayak2017atmospheric}. Due to the assumption of chemical equilibrium, the main absorber in all models shown is CH$_4$. On each subplot we show constant exposure-time contours for detector-noise-limited and photon-noise-limited observations. \textbf{Main Point}: We need SNR$=$20 to \textit{constrain} composition and gravity. Results are less sensitive to resolving power.}
\label{fig:ic}
\end{figure*}
Currently, an important driver for the creation of \texttt{PICASO} is to determine optimal observing strategies for future direct imaging missions, such as \textit{WFIRST}, ELTs, and potential large space based observatories such as LUVOIR or HabEx. For example, determining band pass ranges, minimum SNR, and instrument resolving powers that maximize the total retrievable information from a planetary reflected light spectrum will be a critical contribution to the design of future facilities. Throughout this analysis, we focus specifically on the approximate SNR, bandpass, and resolution of the \textit{WFIRST}-Coronographic Instrument, which is a technology demonstrator for future concept mission like LUVOIR or HabEx. Our methodology, though, can be applied to any parameter space.

\citet{lupu2016developing} and \citet{nayak2017atmospheric} began to explore optimal observing strategies by wrapping the original \texttt{Fortran} code outlined in \citet{cahoy2010exoplanet} and others, in a sophisticated retrieval framework. \citet{lupu2016developing} focused on our ability to ascertain the presence or absence of clouds and CH$_4$, while \citet{nayak2017atmospheric} focused on our ability to constrain planet phase and radius. These studies offered valuable insights into our ability to constrain the atmospheres of exoplanets with reflected light. However, the computational limitations of MCMC (or similar) methods hinders our ability to rapidly move through a large parameter space in atmospheric diversity, resolution, and SNR.

Information content (IC) theory offers an alternative to full MCMC methods. IC has been commonly used in Earth and Solar System science \citep[e.g.][]{kuai2010channel, saitoh2009co2}, as well as in exoplanet science \citep[e.g.][]{line2012info, batalha2017ic,howe2017ic,batalha2018ic}. We use the IC model originally developed for transiting exoplanet science. A full description of the methodology can be found in \citet{batalha2017ic}.

IC theory relies heavily on computing the Jacobian of individual systems, which describe how sensitive the model is to slight perturbations of the state vector parameters at a given initial state. In this analysis we assume that the state vector is made up of [$T(P)$, $\xi_i$, $g$], where $T(P)$ is the pressure-dependent temperature profile, $\xi_i$ is the mixing ratio of species $i$, and $g$ is the gravity. We compute the derivative of the Jacobian using a centered-finite difference scheme. Our $T(P)$ and mixing ratio profiles come from the calculations in \citet{batalha2018color}, so that perturbations shift the entire profile. $T(P)$ and $g$ are perturbed linearly, with 0.1\% perturbations. $\xi_i$'s are perturbed in log space, also with 0.1\% perturbations. These finite perturbations were chosen to reproduce the results of a full retrieval analyses. 

Of course there are several other parameters that contribute to an atmospheric state. Choosing only $T(P)$, $\xi_i$, $g$ is almost certainly too simplistic. For example, as we've seen here, the cloud asymmetry parameter and the single scattering albedo will largely contribute to how well we can constrain the atmospheres of exoplanets. Additionally, \citet{lupu2016developing} showed that the pressure of the cloud deck will also influence the shape of the spectral features. In order to capture this behavior, we compute the Jacobian across a diversity of initial states ($a_s$=0.5-5.0~AU, $f_{\rm sed}$=0.1-6) at a phase angle of 90$^\circ$. As shown in Figure \ref{fig:asy}, this covers a broad diversity of cases in single scattering and asymmetry, to compensate for our simplistic state vector.

Along with the Jacobian, \textbf{K}, we also need an approximation of the error covariance, $\mathbf{S_e}$, and $\mathbf{S_a}$, the \textit{a prior} covariance matrix.  We take the values of the error covariance matrix from the cases in \citet{nayak2017atmospheric} for SNR=5-25. The priors, $\mathbf{S_a}$, represents the information we start with for any given system. We assume broad uniform priors for every state vector parameter. In other words, we assume to have very little information about the system before conducting our observation: $\pm300~K$ for $T(P)$, $\pm6$~dex for mixing ratio, $\pm100$m/s$^2$ for gravity. Given $\mathbf{K}$, $\mathbf{S_e}$, and $\mathbf{S_a}$, we can compute the posterior covariance matrix, which gives the 1-sigma uncertainty on a state vector parameter after a measurement is made:
\begin{equation}
    \mathbf{\hat{S}} = (\mathbf{K^TS_e^{-1}}\mathbf{K} + \mathbf{S_a^{-1}})^{-1}
\end{equation}
Because of the $\mathbf{S_a}^{-1}$ dependence, using large priors guarantees that our estimates for the posterior covariance matrix are solely driven by the model sensitivity (via the Jacobian) and the expected data quality at each wavelength (e.g. $\mathbf{K^TS_e}^{-1}\mathbf{K} >> \mathbf{S_a^{-1}}$ ). Additionally, we spot checked our analysis against the full retrievals done in \citet{lupu2016developing} \& \citet{nayak2017atmospheric}, and found that they are in good agreement. 

Figure \ref{fig:ic} shows a summary of the results of the IC analysis for a subset of $a_s$ and $f_{\rm sed}$. We focus on this subset because it is the ``sweet spot'' in parameter space for a \textit{WFIRST}-CGI mission. However, we discuss the full parameter space in \S\ref{sec:iccomp} \& \S\ref{sec:icgrav}. Additionally, Figure \ref{fig:ic} shows constant exposure time contours for both detector-noise and photon-limited observations. Because \textit{WFIRST} instrumentation has not yet been finalized, we cannot add definitive exposure times on each of these curves. However, we include them anyways to give readers an understanding of the SNR-Resolving Power interplay, in terms of total time (e.g. for detector-noise limited observations it takes equal time to achieve SNR=25 at R=40, as SNR=5 at R=120). As \textit{WFIRST} instrumentation is solidified, we will perform more robust noise simulations with estimates for integration time. 

Overall, our ability to constrain composition and gravity are more dependent on the SNR, as opposed to instrument resolving power. Regardless of resolving power, SNR=10 is not sufficient to \textit{constrain} either composition or gravity (our definition of constraint is discussed in the following \S\ref{sec:iccomp}). Generally, a SNR$\sim$20 is needed to attain robust constraints on composition and gravity. 

\subsection{Sensitivity to Composition}\label{sec:iccomp}
Figure \ref{fig:ic} only shows the ability to constrain the abundance of CH$_4$, the dominant absorber at these temperatures. From $a_s$=0.5-5, there is a transition from alkali-dominated atmospheres (Na \& K) toward 0.5 AU, to CH$_4$-dominated atmospheres toward 5 AU. This transition, a result of chemical equilibrium, roughly occurs at 0.85~AU \citep[see][]{batalha2018color}, where both alkali and CH$_4$ features are comparatively small. \textit{WFIRST}-CGI, with a proposed wavelength coverage for spectroscopy of $\sim 0.6$--$0.76\,\rm \mu m$ will be primarily focused on the detection of CH$_4$. Therefore, we only show figures for $a_s=1$ \& 3~AU because most of the considered targets will fall in this range. 

At SNR=5, constraints on CH$_4$ approach the prior value, meaning the observation does not contribute to the overall knowledge. A definition of ``good'' constraint is relatively arbitrary, but adopt the definition of \citet{feng2018earth}, which is effectively the ability to constrain the abundance within $\pm$1.0 log units. Although, this may seem too stringent a definition, IC analyses tends to be more optimistic than full retrieval analyses. This is because IC cannot pick up on important factors such as degeneracies between state vector parameters. 

The difference in being able to detect CH$_4$ at 1~AU versus 3~AU comes from the relative size of the molecular features, and the cloud composition. At 1~AU, even though Na \& K are nearly gone,  CH$_4$ is still not as pronounced as it is at 3~AU, because the volume mixing ratio is lower. Additionally, higher-in-altitude water clouds will weaken the feature.

The cloud parameter $f_{\rm sed}$ appears to have a lesser effect than semi-major axis. This is because at moderately high values of $f_{\rm sed}$, the cloud deck is at low enough pressures as to not completely impede the detection of molecular features. For $f_{\rm sed}\le1$, detection of CH$_4$ (or any other molecular feature) will be difficult to impossible because the path length for reflected light through the atmosphere is too short.

\subsection{Sensitivity to Gravity}\label{sec:icgrav}
The effect of gravity on reflected light is summarized in Figure 3 \& 4 in \citet{lupu2016developing}. Generally, increasing gravity increases the depth of spectral features and increases reflectively towards the blue. There is also a more subtle effect of gravity on the 0.8~$\mu$m H$_2$ continuum feature (at lower gravity the feature is stronger). We are not able to leverage this effect because of \textit{WFIRST}-CGI's spectroscopic wavelength coverage. 

For SNR$\le 10$ the constraint on gravity, approaches the prior, meaning the observation does yet not contribute to the overall knowledge. It also appears that 1~AU systems are slightly more amenable to gravity characterization as opposed to 3~AU. This because at 3~AU when $f_{\rm sed}\ge1$, water cloud reflectively dominates the opacity, as opposed to at 1~AU, where there is still a contribution from scattering by Rayleigh. When water cloud opacity dominates the opacity, the spectrum is less sensitive to slight perturbations in gravity. This is also why the $f_{\rm sed}$=3 cases are better constrained, as opposed to the $f_{\rm sed}$=1 cases. 

\section{Discussion \& Conclusion}\label{sec:discon}
Here, we presented an initial release of a reflected light code, called \texttt{PICASO}. \texttt{PICASO} is versatile enough for calculations of reflected light spectroscopy, and for retrievals of directly imaged exoplanet atmospheres. It has been benchmarked against two independent codes from \citet{dlugach1974optical} and \citet{madhu2012analytic}. For isotropic and Rayleigh scattering, \texttt{PICASO} agrees with other codes to well within 10\%. For asymmetric scattering, calculations are slightly more discrepant, but well within the bounds of observational precision ($\sim10\%$ agreement). 

\texttt{PICASO} contains several methodologies for computing calculations of reflected light. Specifically, we have focused on highlighting different methodologies for computing single scattering, multiple scattering, and Raman scattering. Within each section, we have provided recommendations for modeling exoplanets, which are also \texttt{PICASO}'s default run settings. A further explanation of this is available in our online radiative transfer tutorial\footnote{\href{https://natashabatalha.github.io/picaso_dev}{\faSearch:Physics Tutorial}}. 

Our information content analysis demonstrates the approximate parameter space in cloud composition, resolving power, and SNR, where we can expect to get robust constraints on composition and gravity. We find that in general, we need an SNR$\sim$20 to attain constraints on composition, where our definition for constraint is attaining a 1-$\sigma$ confidence interval of $\pm$1 log unit on the volume mixing ratio of the dominant absorber \citep[following][]{feng2018earth}. 

Despite the versatility of the original release, there are still aspects which we are currently working on. Future releases of the code will contain: 
\begin{enumerate}
    \item Thermal Emission 
    \item $\delta$-M stream method \citep{wiscombe1977deltam}
    \item Raman scattering by N$_2$ and He
    \item Chebyshev polynomial for multiple scattering phase function
    \item Compatibility with nested sampling algorithm
\end{enumerate}
Additionally, a robust retrieval analysis will be needed to address degeneracies that cannot be captured in an information content analysis. This includes developing methods to constrain radius, directly retrieve the optical properties of the clouds (i.e. the imaginary component of the refractive index), and discern the presence of photochemical hazes. This analysis will additionally be added as a future code release since \texttt{PICASO} contains the modularity and versatility to support it.

\acknowledgments
We thank Kerri Cahoy for helpful discussion and tracking down various versions of the original albedo code. Additionally, we thank Cornell undergraduate Mark Siebert, and Caltech graduate student Danica Adams for being the first beta testers, and pointing out some bugs in the code and installation. N.E.B acknowledges support from the University of California President’s Postdoctoral Fellowship Program. M.S.M. acknowledges support from GSFC Sellers Exoplanet Environments Collaboration (SEEC), with funding specifically by the NASA Astrophysics Division’s Internal Scientist Funding Model.

\software{numba \citep{numba}, pandas \citep{mckinney2010data}, bokeh \citep{bokeh}, NumPy \citep{walt2011numpy}, IPython \citep{perez2007ipython}, Jupyter, \citep{kluyver2016jupyter},PySynphot \citep{pysynphot2013}, sqlite3 \citep{sqlite3},picaso\citep{picaso} }

\bibliographystyle{aasjournal}
\bibliography{picaso}

\appendix
\section{List of All Modeling Recommendations}
In \S\ref{sec:datass} we explored \texttt{PICASO}'s sensitivities to single scattering phase function, multiple scattering phase function, and Raman scattering methodology. Throughout the text, we outlined our modeling suggestions. Here, we aggregate those recommendations into a single table. In this version of \texttt{PICASO}, these represent the current radiative transfer defaults. 

\begin{itemize}
    \item \textbf{Single Scattering:}
            \begin{itemize}
                \item Use default specification for direct scattering (\texttt{TTHG\_Ray}) 
                \item Fit for the functional form of the fraction, $f$, of forward to back scattering according to the problem being addressed. 
                \item See Table \ref{tab:single} for a list of pros/cons
            \end{itemize}
    \item \textbf{Multiple Scattering:}
    \begin{itemize}
        \item For planet cases with some degree of asymmetric cloud scatterers, always use N=2 Legendre polynomial expansion with the $\delta$-Eddington correction.
    \end{itemize}
    \item \textbf{Raman Scattering:}
    \begin{itemize}
        \item Choose Pollack et al. methodology with \citet{antonija2016raman} cross sections. 
        \item Choose a stellar spectrum that matches the level of resolution needed
        \item The user will experience slight compute speed losses for a single run depending largely on the stellar/planet resolution chosen. However, because these shifts only need to be computed once, adding this Raman scattering methodology is not a computational burden. 
    \end{itemize}
    \item \textbf{Phase Geometry}:
    \begin{itemize}
        \item The default number of integration angles is 10 Gauss and 10 Chebyshev angles. If the user is particularly interested in exploring scattering effects at high cosine angle (e.g. near planet limb), it would be beneficial to increase the number of planet facets despite the decrease in computation speed.
    \end{itemize}
\end{itemize}

\section{Opacity Database}
For this version of the code, \texttt{PICASO} contains a database of opacities that are hosted on \texttt{Github}. As summarized in \citet{freedman2008opacities}, our molecular opacities are computed on a 1060 point pressure-temperature grid from 0.3-1$\mu$m. Currently this database contains molecular opacity from: CH$_4$, CO$_2$, CrH, FeH, H$_2$O, H$_2$S, K, Li, NH$_3$, Na, Rb, TiO, and VO. Notably, for CH$_4$ we include the visible methane following \citet{karkoschka1994spec}. For continuum absorption, we include H bound-free, H free-free, H$_2^-$, H$_2$-CH$_4$, H$_2$-H, H$_2$-H$_2$, H$_2$-He, H$_2$-N$_2$. We also include Rayleigh scattering from H$_2$, He, and CH$_4$, and Raman scattering from H$_2$ \citep{antonija2016raman}. 

Our opacity database is constructed in \texttt{sqlite3} format. \texttt{sqlite3} is a user-friendly, python-based module for the \texttt{C}-library, \texttt{SQLite}. After testing several database formats (\texttt{json}, \texttt{hdf5}, \texttt{ascii}, \texttt{sqlalchemy}), \texttt{SQLite} was chosen because it is a lightweight disk-based database that does not require a separate server process. Additionally, as we expand our opacity database it will be trivial to port over this smaller \texttt{SQLite} database to another format that can handle much larger data structures. We provide a full tutorial on how to query and construct \texttt{sqlite3} databases\footnote{\href{https://natashabatalha.github.io/picaso/notebooks/5_SwappingOpacities.html}{\faCode:Opacity Tutorial}}. If users follow our recipe, they can swap in any molecular opacities without needing any code modifications.



\end{document}